\documentclass[twocolumn,superscriptaddress,floatfix,preprintnumbers,amssymb ,amsmath]{revtex4}
\usepackage{graphicx}
\usepackage{dcolumn}
\usepackage{bm}
\usepackage[latin1]{inputenc}
\usepackage[mathscr]{eucal}
\usepackage{epsfig}
\usepackage{rotating}

\begin{document}

\title{Radio-Frequency Single-Electron Refrigerator}

\author{Jukka P. Pekola}
\affiliation{Low Temperature Laboratory, Helsinki University of
Technology, P.O. Box 3500, 02015 TKK, Finland}

\author{Francesco Giazotto}
\affiliation{NEST CNR-INFM \& Scuola Normale Superiore, I-56126 Pisa,
Italy}

\author{Olli-Pentti Saira}
\affiliation{Low Temperature Laboratory, Helsinki University of
Technology, P.O. Box 3500, 02015 TKK, Finland}


\begin{abstract}
We propose a cyclic refrigeration principle based on mesoscopic
electron transport. Synchronous sequential tunnelling of electrons
in a Coulomb-blockaded device, a normal metal-superconductor
single-electron box, results in a cooling power of $\sim k_{\rm B}T
\times f$ at temperature $T$ over a wide range of cycle frequencies
$f$. Electrostatic work, done by the gate voltage source, removes
heat from the Coulomb island with an efficiency of $\sim k_{\rm
B}T/\Delta$, where $\Delta$ is the superconducting gap. The
performance is not affected significantly by non-idealities, for
instance by offset charges. We propose ways of characterizing the
system and of its practical implementation.
\end{abstract}

\maketitle Cyclic conversion of heat into mechanical work, or
conversely, work into extracted heat, demonstrates basic concepts of
thermodynamics. The latter forms the basis of refrigeration (see,
e.g., \cite{reichl}). Whereas cryogenic refrigeration dates back to
the 19th century \cite{reichl,pobell}, electronic microrefrigeration
has existed only for a little longer than a decade
\cite{giazotto06}. Successful solid-state refrigeration schemes
operating at cryogenic temperatures, typically in the sub-kelvin
regime, rely most commonly on normal metal-insulator-superconductor
(NIS) tunnel junctions \cite{giazotto06,nahum}. They have allowed
substantial electron \cite{jp04,leivo} as well as phonon
\cite{luukanen,clark} temperature reduction. Quasiparticle
refrigeration occurs thanks to the superconducting energy gap
which allows only most energetic electrons to escape in a tunneling
process, thus effectively cooling the N region. Until now no
electronic cyclic refrigerators have, to the best of our knowledge,
been demonstrated. Yet there exists a proposal to use a mesoscopic
semiconductor ratchet as a Brownian heat engine for electrons
\cite{humphrey02}. In this Letter we show that a simple mesoscopic
hybrid nano-structure in the Coulomb blockade regime exhibits
significant cooling power when work is done on it by a
radio-frequency gate, even in the absence of DC-bias. The presented
concept bears resemblance to that of a single-electron pump, where
an electron is transported synchronously at an operation frequency
$f$ through a device producing average electric current of magnitude
$ef$ \cite{pothier92,keller96,keller99}. In the present system, heat
at temperature $T$ is transported similarly due to cyclic gate
operation in the Coulomb blockade regime with average "current" of
order $k_{\rm B}Tf$.
\begin{figure}
    \begin{center}
    \includegraphics[width=8cm]{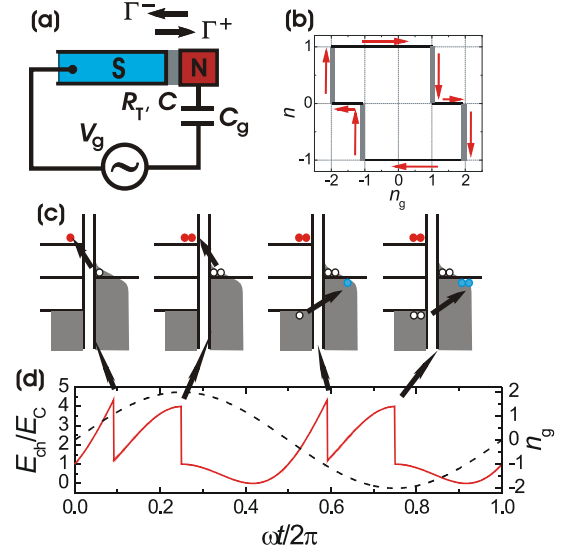}
    \end{center}
    \caption{(Color online) Single-electron refrigerator (SER). (a) Single-electron box with a normal metal (N) island and a
    superconducting (S) lead. (b) The trajectory on the $(n,n_{\rm g})$ plane for
    a representative
    cycle studied. (c) Sketch of energy band diagrams of the SER showing the
    tunneling processes in this cycle. (d) The charging energy
    of the system (solid line, left scale), where discontinuities are observed as electrons tunnel. This
    change of energy, provided by the voltage source, is supplied to
    the tunneling electrons to overcome the superconducting gap.
    The gate cycle is shown
    by the dashed line (right scale).}
    \label{fig:devices}
\end{figure}

As a basic but representative example we study a single-electron box
(SEB) with a normal metal island and a superconducting lead [see
Fig. \ref{fig:devices} (a)], a system whose electrical properties
are determined by single-electron charging effects \cite{SET}. In
practise it can be easily realized with present-day fabrication
technologies.
In SEB, a tunnel junction of capacitance $C$ and resistance $R_{\rm
T}$ connects the island to the lead. A source providing voltage
$V_{\rm g}$ is furthermore capacitively coupled to the island
through a capacitance $C_{\rm g}$.
The charging energy of the
device can be written as
$E_{\rm ch}(n,n_{\rm g})= E_{\rm C}(n+n_{\rm g})^2$.
Here, $E_{\rm C}=e^2/(2C_{\Sigma})$ is the unit of charging energy,
$n$ is the number of extra electrons on the island, and $n_{\rm
g}=C_{\rm g}V_{\rm g}/e$ is the amount of charge in units of $e$
induced by the gate on the island. The total capacitance of the
island is $C_{\Sigma}=C +C_{\rm g}+C_0$, where $C_0$ is the
self-capacitance of the N region.

First we study charge and energy transport in the device, which
allows us to obtain the main result of this paper, i.e., heat
extraction in a cyclic operation of the gate voltage. In the
following this system will be referred to as \emph{single electron
refrigerator} (SER). We identify two stochastic events, tunneling
into the island from the lead, which we associate with a "$+$"
index, and tunneling out from the island to the lead ("$-$" index).
The respective changes in charging energy read:
\begin{eqnarray} \label{dEch}
&& \epsilon^+ \equiv \delta E_{\rm ch}^+/\Delta =2\epsilon_{\rm
C}(n+n_{\rm g}+1/2)\nonumber \\ &&\epsilon^- \equiv \delta E_{\rm
ch}^-/\Delta =-2\epsilon_{\rm C}(n+n_{\rm g}-1/2).
\end{eqnarray}
Here, $\Delta$ is the superconducting energy gap and $\epsilon_{\rm
C}\equiv E_{\rm C}/\Delta$.

Within the theory of sequential single-electron tunneling
\cite{Averin1}, assuming that the tunneling electron does not
exchange energy with the environment \cite{SET}, the two rates of electron
transport can be written as
\begin{eqnarray} \label{Gammas}
&& \Gamma^+(\epsilon^+) =\frac{\Delta}{e^2R_{\rm T}}\int d\epsilon
\tilde{n}_{\rm
S}(\epsilon)\tilde{f}_{\text{S}}(\epsilon)[1-\tilde{f}_{\text{N}}(\epsilon-\epsilon^+)] \nonumber \\
&& \Gamma^-(\epsilon^-)=\frac{\Delta}{e^2R_{\rm T}}\int d\epsilon
\tilde{n}_{\rm
S}(\epsilon)\tilde{f}_{\text{N}}(\epsilon+\epsilon^-)[1-\tilde{f}_{\text{S}}(\epsilon)].
\end{eqnarray}
Here, the normalized BCS density of states at energy $E\equiv
\epsilon \Delta$ reads $\tilde{n}_{\rm
S}(\epsilon)=\epsilon/\sqrt{\epsilon^2 -1}$. To take into account
non-idealities in realistic superconductors \cite{dynes84}, we use
$\tilde{n}_{\rm S}(\epsilon) = |\rm{Re}[(\epsilon +
i\gamma)/\sqrt{(\epsilon + i\gamma)^2-1}]|$ instead, where the
typical value of the smearing parameter $\gamma$ is in the range
$1\cdot 10^{-5}...1\cdot 10^{-3}$ for aluminum as a thin-film
superconductor \cite{jp04}. Furthermore,
$\tilde{f}_{\text{S(N)}}(\epsilon)$ is the Fermi-Dirac distribution
function in S (N) at temperature $T_{\text{S}}(T)$ and energy $E$.
The corresponding heat fluxes associated with the two tunneling
processes are given by
\begin{eqnarray} \label{qs}
\dot{Q}^+ (\epsilon^+) =\frac{\Delta^2}{e^2R_{\rm T}}\int d\epsilon
(\epsilon - \epsilon^+) \tilde{n}_{\rm
S}(\epsilon)\tilde{f}_{\text{S}}(\epsilon)[1-\tilde{f}_{\text{N}}(\epsilon-\epsilon^+)]
\nonumber \\ \dot{Q}^- (\epsilon^-) =\frac{\Delta^2}{e^2R_{\rm
T}}\int d\epsilon (\epsilon + \epsilon^-) \tilde{n}_{\rm
S}(\epsilon)\tilde{f}_{\text{N}}(\epsilon+\epsilon^-)[1-\tilde{f}_{\text{S}}(\epsilon)],
\end{eqnarray}
where the first one is the energy deposition rate by incoming
electrons, and the second one is the energy extraction rate by
leaving electrons.

We start by an approximate calculation which yields
illustrative yet nearly quantitative results.
At low temperatures, $T,T_{\text{S}}\ll\Delta/k_{\rm B}$, and in the
domain $1+\epsilon^\pm > 0$, Eqs. \eqref{Gammas} yield
\begin{equation} \label{gm}
\Gamma^\pm (\epsilon^\pm)\simeq \frac{1}{e^2R_{\rm
T}}\sqrt{\frac{\pi}{2}\Delta k_{\rm B}T}\,\,\,
e^{-\frac{\Delta}{k_{\rm B}T}(1+\epsilon^\pm)}.
\end{equation}
With the same approximations, we get from Eqs. \eqref{qs}
\begin{equation} \label{qm}
\langle Q^\pm (\epsilon^\pm)\rangle \simeq \mp[\frac{k_{\rm B}T}{2}
+ \Delta (1+\epsilon^\pm)]
\end{equation}
for the average energy per event of a tunneling electron with
respect to the Fermi energy, $\langle Q^\pm (\epsilon^\pm) \rangle
\equiv \dot{Q}^\pm (\epsilon^\pm)/\Gamma^\pm (\epsilon^\pm)$. We
expect the exponent in Eq. \eqref{gm} to assume values in the range
1...10 at that point in the cycle where tunneling is most likely to
occur. This is because the prefactor in front of the exponential
term, the "attempt frequency", is typically a few orders of
magnitude higher than the operation frequency of the device
\cite{parameters}. The tunneling electron is then expected to have
energy
\begin{equation} \label{qm}
Q^\pm \sim \mp k_{\rm B}T.
\end{equation}
This states that the "$+$" process brings a \emph{negative} amount
of energy into the island, whereas the "$-$" process removes a
\emph{positive} energy from the island, i.e., all the events tend to
extract hot excitations from the N island, which results in
refrigeration of it. The expectation value of energy carried out
from the island in a full cycle is then
$Q  = n_{\rm t}( Q^-  -  Q^+ )$,
where $n_{\rm t}$ is the number of tunneling events in both
directions within a full cycle. We then have
$Q = rk_{\rm B}T$,
where $r$ is of the order of $n_{\rm t}$. By repeating cycles at frequency
$f$ one obtains an average cooling power
\begin{equation} \label{coolingp}
\dot{Q} = rk_{\rm B}Tf.
\end{equation}
To go beyond these approximate results, we employ either
\emph{deterministic}, master equation-based computations, or
\emph{stochastic} simulations of charge and energy dynamics.
\begin{figure}[t!]
    \begin{center}
    \includegraphics[width=8cm,clip]{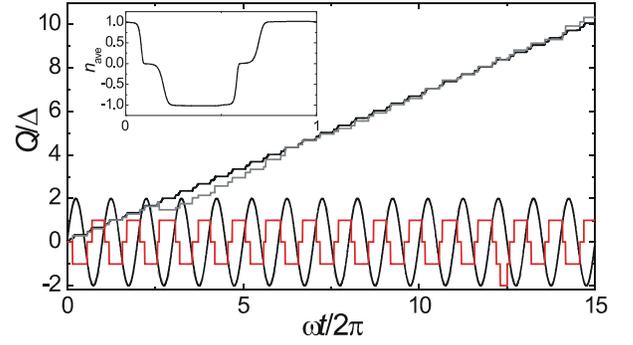}
    \end{center}
    \caption{(Color online) Heat extracted from the N island
    in a SER during the cyclic operation of the gate.
    These
    calculations were performed with the following parameter values:
    $k_{\rm B}T/\Delta =0.05$, $\epsilon_{\rm
C}\equiv E_{\rm C}/\Delta = 0.3$, $R_{\rm T} = 30 $ k$\Omega$,
$\gamma =1\cdot 10^{-4}$, $f \equiv \omega/(2\pi) = 10$ MHz, and
$\Delta = 200$ $\mu$eV, which corresponds to aluminum as the
superconductor.
    The gray line shows the
    the stochastic results, and the rising black line is the solution of the master equation.
    Sinusoidal line indicates the instantaneous gate position $n_{\rm g}$, and the piece-wise constant red line
    shows the number of extra electrons on the island in a typical trajectory.
    The inset displays the average number of charges, $n_{\rm ave}$,
    at each instant within the cycle obtained from the master equation.}
    \label{fig:results}
\end{figure}
In the first method, we follow the probabilities $p_n(t)$ of
observing charge number $n$ on the island at time instant $t$.
Equations~(\ref{Gammas}) yield for each $n$ state the time-dependent
transition rates $\Gamma^\pm_n$ corresponding to the "$+$" and "$-$"
processes. The master equation then reads
\begin{equation}
\dot{p}_n = -(\Gamma^+_n + \Gamma^-_n) p_n + \Gamma^+_{n-1} p_{n-1}
+ \Gamma^-_{n+1} p_{n+1}. \label{master}
\end{equation}
Furthermore, we write the expectation value of instantaneous heat
current with the help of Eqs.~(\ref{qs}) as
\begin{equation}
P = \sum_n p_n (\dot{Q}^-_n - \dot{Q}^+_n). \label{master_Qdot}
\end{equation}
We keep track of only a small number ($\sim k_{\rm B} T
 / E_{\rm C}$) of lowest-energy charge number states, and assume the rest
to have zero occupation probability. With a reasonable initial
condition, e.g., $p_n(0) = \delta_{n 0}$, $p_n(t)$ converges to an
approximate periodic equilibrium during the first few cycles. The
total cooling power $\dot{Q}$ is then obtained by averaging
Eq.~(\ref{master_Qdot}) over one cycle.

In the stochastic method, we employ a Monte Carlo simulation of the
tunnel events to generate a sample trajectory in $n$ space. We
evaluate the integrands in Eqs. \eqref{Gammas} over a small energy
interval $\delta\epsilon$. Multiplied by a small time step $\delta
t$, these yield the probability of observing a "$+$" event at energy
$\epsilon - \epsilon^+$ or a "$-$" event at energy $\epsilon +
\epsilon^-$ during $\delta t$. If the "$-$" ("$+$") process occurs,
the energy is added to (subtracted from) the cooling budget. The
total cooling power can be obtained by averaging over a trajectory
long enough to suppress the stochastic variation to a negligible
level.



As an example we consider a cycle where the gate voltage varies
periodically such that $n_{\rm g}=2\sin(\omega t)$, see Fig.
\ref{fig:devices} (d). As presented in Fig. \ref{fig:devices} (b),
$n$ then tends to follow the sequence $+1 \rightarrow 0 \rightarrow
-1 \rightarrow 0 \rightarrow +1$. The thick vertical lines indicate
the stochastic uncertainty of the tunnel events along the gate
excursion. In Fig. \ref{fig:devices} (c) and (d) we show,
respectively, the tunnel events on an energy diagram during the
example cycle demonstrating the cooling feature of the device, and
the corresponding charging energy along the cycle. Figure
\ref{fig:results} shows numerical results over 15 cycles, of charges
on the island and heat extracted from it. In the stochastic
simulation the ideal cycle of $n: +1 \rightarrow 0 \rightarrow -1
\rightarrow 0 \rightarrow +1$ is repeated almost regularly over
these cycles, although the system visits once the $n=-2$ state and
the exact time instants of tunneling jitter around their mean
values.
The deterministic calculation yields also the expectation value of
charge on the island at each value of gate (inset). The efficiency
of the refrigerator, defined as the ratio of heat extracted and work
done by the source, is $\eta \simeq 4.0 k_{\rm B}T/\Delta$ in this
particular case.
This exceeds by more than a factor of three the corresponding figure
of a static NIS refrigerator, $\eta \simeq 1.2 k_{\rm B}T/\Delta$
\cite{giazotto06}. This difference stems from yet another favourable
energy filtering in SER: due to Coulomb blockade, only one electron
tunnels in each phase of the cycle, and it extracts maximal amount
of heat from N.

\begin{figure}[t!]
    \begin{center}
    \includegraphics[width=8cm]{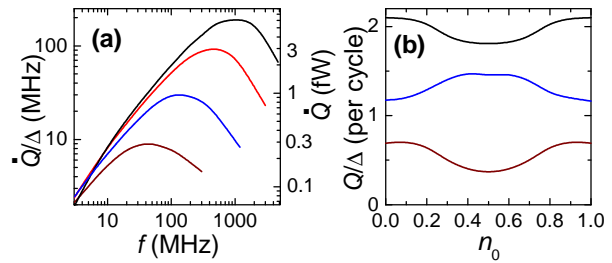}
    \end{center}
    \caption{(Color online) (a) Heat flux extracted from the N island of
    a SER as a function of frequency for some values of $R_{\rm T}$.
    From bottom to top: $R_{\rm T}=100$, 30, 10, and 5 k$\Omega$.
    The computations were performed assuming
    $n_{\rm g}(t)= 2 \sin(\omega t)$. (b) Normalized energy extracted per cycle versus
    background charge $n_0$.
    In the calculations we used $f=10$ MHz, and $R_{\rm T}=30$ k$\Omega$.
    The gate amplitudes for the three curves from bottom to top are 2, 3.5, and
    5. In both (a) and (b) we assumed $\Delta=200$ $\mu$eV, $k_{\rm B}T/\Delta =0.05$, $E_{\rm C}/\Delta =0.3$, and
    $\gamma = 1\cdot 10^{-4}$.}
    \label{fig:frequency}
\end{figure}
Further insight into the performance of SER can be gained by
inspecting the frequency dependence of $\dot{Q}$, shown in Fig.
\ref{fig:frequency} (a) for different values of $R_{\text{T}}$.
$\dot{Q}$ is a non-monotonic function of $f$, decreasing both at low
and high values of $f$. Increasing the junction transparency leads
to an enhancement of the maximum heat current and to widening of the
window of operational frequencies. This can be easily understood in
terms of reduced discharging time through the junction. For
$R_{\text{T}}=30$ k$\Omega$ (5 k$\Omega$), $\dot{Q}$ is maximized
around $f\simeq 150$ MHz (1 GHz), where it obtains values of $\sim
1$ fW ($\sim 6$ fW) for Al. As will be shown below, such values will
lead to substantial electronic temperature reduction in the N
island.

Whether a single-electron device is usable in practice depends on
its immunity to fluctuating background charges ($Q_0$). Their
influence can be studied numerically by adding an offset $n_0\equiv
Q_0/e$ to $n_{\rm g}$.
The results are displayed in Fig. \ref{fig:frequency} (b), which
indicates heat extracted per cycle vs $n_0$ for three values of gate
amplitude. Clearly $n_0$ has some effect on the cooling power of
SER, but these small variations can be compensated for by adjusting
either the amplitude or the DC-bias of the gate, if necessary.
Furthermore, in practice it may be advantageous to suppress the
influence of $n_0$ by employing larger gate amplitudes at lower $f$,
which leads to the same cooling power, but to weaker $n_0$
dependence.

In order to test the performance of the SER in a realistic context,
we need to take into account the mechanisms that drive power into
the N island. We expect the system to be in the quasi-equilibrium
limit \cite{giazotto06}, where strong inelastic electron-electron
interaction forces the quasiparticle distribution toward a thermal
one at a temperature $T$ which can differ from that of the lattice
($T_{\text{bath}}$). At low temperatures, which for metallic
structures are typically below 1 K, the main heat load on electrons
comes from phonon scattering with energy exchange according to
$\dot{Q}_{\text{e-bath}}(T,T_{\rm bath})=\Sigma
\mathcal{V}(T^5-T_{\text{bath}}^5)$ \cite{roukes85}, where $\Sigma$
is the electron-phonon coupling constant and $\mathcal{V}$ is the
island volume. In order to cool the island, frequency $f$ has to be
high enough to prevent full relaxation toward $T_{\text{bath}}$.
Thus we require $(\tau_{\text{e-ph}})^{-1}\ll f \ll
(\tau_{\text{e-e}})^{-1}$, where $\tau_{\text{e-ph(e-e)}}$ is the
electron-phonon (electron-electron) inelastic scattering time. These
constraints can be met experimentally recalling that, e.g., for a
copper (Cu) island one has $(\tau_{\text{e-ph}})^{-1}<10^5$ Hz at
$T\lesssim$ 200 mK \cite{taskinen}, while
$(\tau_{\text{e-e}})^{-1}\sim10^9-10^{10}$ Hz even in quite pure
metal samples \cite{pierre03}. Heat input by other relaxation
mechanisms, in particular radiative electron-photon heat load, can
be ignored in this case because of poor matching
\cite{schmidt04,meschke06}.

\begin{figure}[t!]
    \begin{center}
    \includegraphics[width=8cm]{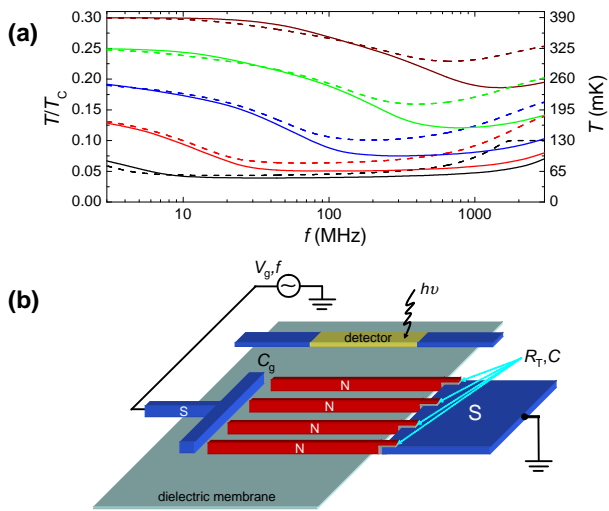}
    \end{center}
    \caption{(Color online)
    (a) Temperature of
    the N island vs gate frequency at a few bath temperatures.
    We assume that the volume of the island is
    $\mathcal{V} = 1\cdot 10^{-21}$ m$^3$, $\Sigma = 1\cdot 10^{9}$ WK$^{-5}$m$^{-3}$,
    $E_{\rm C}/\Delta =0.3$, and $\gamma = 1\cdot 10^{-4}$. $T_{\text{C}}=\Delta/(1.764k_{\text{B}})$ is
    the superconducting critical temperature.
    The results are shown for $R_{\rm T} = 10$ k$\Omega$ (solid lines) and $R_{\rm T} = 30$ k$\Omega$ (dashed lines), and
    for bath temperatures (from bottom to top) 131 mK ($T_{\text{bath}}/T_{\text{C}}=0.1$, black), 197 mK ($T_{\text{bath}}/T_{\text{C}}=0.15$, red),
    263 mK ($T_{\text{bath}}/T_{\text{C}}=0.2$, blue), 328 mK ($T_{\text{bath}}/T_{\text{C}}=0.25$, green) and
    394 mK ($T_{\text{bath}}/T_{\text{C}}=0.3$, brown). (b) A scheme for
    refrigerating
    a device on a thin dielectric platform using an array of SERs.
    The SERs cool down electrons and phonons on the membrane, as well as the device on the platform.
    The islands share a common gate and a superconducting lead outside the
    membrane.}
    \label{fig:performance}
\end{figure}
In Fig. 4 (a) we plot the steady state temperature of an N island of
a typical size. The results are shown as a function of $f$ at few
representative values of $T_{\rm bath}$, which we furthermore assume
to be equal to $T_{\rm S}$. The lines are solutions of the balance
equation $\dot{Q}(T) + \dot{Q}_{\text{e-bath}}(T, T_{\rm bath}) =
0$, where $\dot{Q}(T)$ is the cooling power of SER at temperature
$T$. The results demonstrate significant temperature reduction over
a broad range of frequencies. In practise $T$ can be measured, e.g.,
by an additional NIS junction, which probes the temperature
dependence of a tiny quasiparticle current. A thermometer based on
supercurrent in a Coulomb-blockaded SINIS junction is another
possibility \cite{ostrovsky04}. In Fig. 4 (b) we show a possible
implementation of an array of SERs to refrigerate a separate device,
e.g., a radiation detector on a dielectric membrane \cite{clark}. In
this realization, the performance of SERs can be probed by measuring
the drop of lattice temperature \cite{luukanen}.

We presented the concept and analyzed the performance of a cyclic
single-electron refrigerator, which introduces thermodynamic cycles
in a nano-electronic system. The concept can readily be generalized
to more complex structures with further enhanced performance. For
example, a three terminal device, in form of a SINIS single-electron
transistor, can be controlled additionally by applying a
drain-to-source voltage across. Coulomb blockade suppresses the
cooling power of a statically biased SINIS refrigerator
\cite{unpublished}, but adding radio-frequency gating can enhance
the cooling power even above that of the SER presented here.
Finally, we analyzed here quantitatively only sinusoidal cycles of
$n_{\rm g}$ in the basic SER. Alternative architectures and cycles
may lead to better control between $\dot{Q}$ and $k_{\rm B}Tf$, and
the device might then yield a "quantized" and synchronized heat
current in analogy to electrical current in metrological devices.

We thank Mikko Paalanen for helpful discussions and Academy of
Finland (TULE) and the EC-funded ULTI Project, Transnational Access
in Programme FP6 (Contract \#RITA-CT-2003-505313) for financial
support.


\begin{thebibliography}{99}
\bibitem{parameters} We may assume $k_{\rm B}T/\Delta \sim 0.1$
for temperature to be well below the superconducting transition
temperature $T_{\rm C}$ ($\simeq 1$ K for a usual superconductor
such as Al), and $R_{\rm T} \simeq h/e^2 \simeq 26 $ k$\Omega$ for
junction to be well in the Coulomb blockade regime. Then
$\frac{1}{e^2R_{\rm T}}\sqrt{\frac{\pi}{2}\Delta k_{\rm B}T} \sim $
10 GHz, which is well above the operation frequencies that we
envision here, say 1 MHz...1 GHz. Thereby our argument is quite
generally valid.
\end{thebibliography}
\end{document}